\documentclass[12pt]{iopart}
\usepackage{graphicx}
\usepackage{iopams}
\begin{document}

\title{High-energy-resolution molecular beams for cold collision studies}
\author{L.~P. Parazzoli, N. Fitch, D.~S. Lobser, and H.~J. Lewandowski}

\address{JILA, National Institute of Standards and Technology and\\ University of Colorado, and Department of Physics,
University of Colorado,\\ Boulder, Colorado 80309-0440}
\ead{parazzoli@colorado.edu}
\begin{abstract}
Stark deceleration allows for precise control over the velocity of a pulsed molecular beam and, by the nature of its limited phase-space acceptance, reduces the energy width of the decelerated packet. We describe an alternate method of operating a Stark decelerator that further reduces  the energy spread over the standard method of operation.  In this alternate mode of operation, we aggressively decelerate the molecular packet using a high phase angle.  This technique brings the molecular packet to the desired velocity before it reaches the end of the decelerator; the remaining stages are then used to longitudinally and transversely guide the packet to the detection/interaction region.  The result of the initial aggressive slowing is a reduction in the phase-space acceptance of the decelerator and thus a narrowing of the velocity spread of the molecular packet.  In addition to the narrower energy spread, this method also results in a velocity spread that is nearly independent of the final velocity. Using the alternate deceleration technique, the energy resolution of molecular collision measurements can be improved considerably.

\end{abstract}

\maketitle

\section{Introduction}

\par
Cold molecular beams created from supersonic jets have been used extensively to study gas-phase reaction dynamics and molecular interactions \cite{Scoles1988, macdonald1989,skodje2000}. Crossed molecular beams are responsible for a large part of our understanding of bimolecular gas-phase reactions. In these systems, molecular beams are created with temperatures as low as 1K with only a few quantum states occupied. This control over internal and external degrees of freedom is necessary to explore the detailed nature of molecular collisions. One parameter that has not been under precise control until recently is the mean velocity of the beam and thus the collision energy. Typical molecular beams can be produced with mean speeds ranging from 300--2000 m/s in large incremental steps depending on the carrier gas used in the expansion. This coarse speed control does not allow experimental studies of narrow collision resonances and thresholds \cite{EricR.Hudson2006,ManuelLara2007}. More recently, crossed-molecular-beam experiments were built in which the relative angle between the beams could be adjusted to allow the collision energy to be changed \cite{GregoryHall1983,sonnenfroh1991}.  These experiments, however, have an energy resolution that is limited by the velocity spread in the initial beam, which is particularly poor at small collision energies.
\par
With the advent of the Stark decelerator, we can now continuously tune the speed of a pulsed molecular beam using time-varying inhomogeneous electric fields \cite{HendrickL.Bethlem2006}. In addition, the nature of the Stark decelerator allows the energy resolution of the beam to be increased greatly. This control enables a new range of molecular collision experiments \cite{MeijerScience2006,Sawyer2008}. Many groups around the world have built experiments based around a Stark decelerator, all of which have the same basic configuration \cite{HendrickL.Bethlem2006,J.R.Bochinski2003,Bucicov2008}. In addition to the standard configuration, which we use for this paper, there are also optical and magnetic analogs \cite{barker2002,coilGun2007} as well as alternating gradient decelerators for decelerating high-field-seeking states  \cite{Bethlem2002,Tarbutt2004} as well as low-field seeking states \cite{Wohlfart2008}.  We describe an alternate mode of operation of a typical decelerator that optimizes the energy resolution for molecular collision experiments.

\section{Standard Stark deceleration}
\par
The Stark deceleration process uses the interaction of an electric field with a molecule's dipole moment to decelerate a portion of a molecular beam. The pulsed beam of molecules is prepared via supersonic expansion in the ground ro-vibrational state in either a ground or metastable electronic state. After the beam is expanded fully, the molecules propagate into a region of the vacuum system containing a series of high-voltage electrode pairs. The geometry of the electrodes creates a maximum of the electric field in the longitudinal direction directly between an electrode pair. The molecules slated for deceleration are in a quantum state that increases in energy with increasing electric field. As these molecules propagate into the increasing electric field, longitudinal kinetic energy is converted to potential energy. If the molecules were allowed to continue down the potential hill, they would regain the lost kinetic energy as they exited the high electric field; however, before they begin to accelerate, the electric field is turned off nearly instantaneously ($<$ 100 ns), thus removing energy from the molecules. We repeat this process with successive stages of electrodes until the molecules have been slowed to the desired speed.  Transverse guidance of the molecules is achieved because the molecules are attracted to the minimum of the electric field along the center of the decelerator. Successive electrode pairs are orientated orthogonally to one another to guide the molecules equally in both transverse dimensions.
\begin{figure}
\center
\includegraphics[scale=.7]{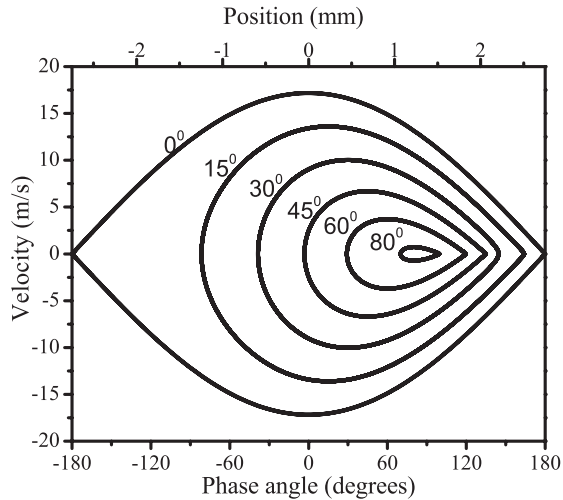}
\caption{ Separatrix for the ND$_3$ molecule in our Stark decelerator. The separatrix curve shows the boundary between the stable and the unstable phase space for a variety of different slowing angles. Molecules within the separatrix curve will be slowed and transported to the end of the decelerator. The separatrix only takes into account the longitudinal dimension of the phase space.}
\label{separatrix}
\end{figure}
\par
The final velocity of the molecular packet is determined by the amount of energy removed per slowing stage and the number of stages. The farther up the potential hill the molecules travel before the field is switched, the more energy is removed from the packet. It is typical to parameterize the position of the synchronous molecule when the field is switched in terms of a phase angle, $\phi_0$ \cite{BethlemHL2002}. A phase angle of $\phi_0=0^{\circ}$ corresponds to no energy being removed from the packet, whereas a phase angle of $\phi_0 = 90^{\circ}$  corresponds to the maximum possible energy being removed. For a particular phase angle, there exists a phase-space acceptance of the decelerator in both position and velocity. The acceptance of the decelerator is the phase-space volume that will be stably slowed and transported to the end of the decelerator. This phase-stable volume can be illustrated through the use of a separatrix (figure \ref{separatrix}). Figure \ref{separatrix} shows the phase-stable portion of the molecular beam decreases rapidly with increasing phase angle. Although large phase-angle slowing reduces the total number of molecules in the packet, it also reduces the velocity spread. We take advantage of this reduced velocity spread in our alternate slowing protocol to produce tunable molecular beams with narrow energy spreads for collision studies.

\begin{figure}
\center
\includegraphics[scale=0.5]{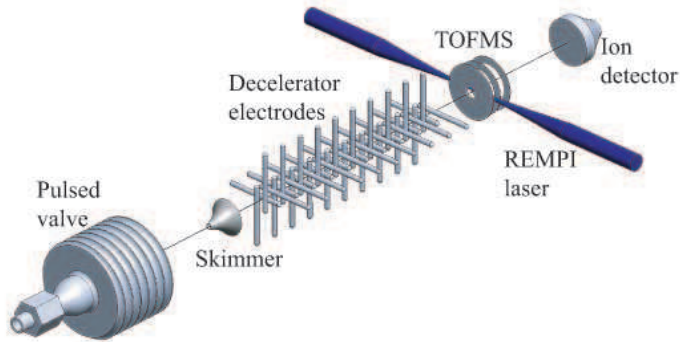}
\caption{ Experimental set-up. It consists of a PZT-driven pulsed valve, molecular beam skimmer, decelerator stages, linear time-of-flight mass spectrometer (TOFMS), and a microchannel-plate ion detector.  The decelerator consists of 150 electrode pairs (not all are shown).}
\label{apparatus}
\end{figure}
\par
We demonstrate the new slowing method with our experimental apparatus shown in figure \ref{apparatus}. It includes a pulsed valve that produces a beam of ND$_3$ molecules seeded in krypton. The mean speed of the beam is  415 m/s with a longitudinal velocity width of 28$\%$. After the expansion, the beam enters a differentially pumped chamber containing a 149-stage Stark decelerator operated at $\pm$ 12 kV. At the exit of the decelerator, the molecules are ionized using a focused beam from a pulsed dye laser operating around 317 nm via a 2+1 resonance enhanced multi-photon ionization (REMPI) process. The ions are then accelerated towards a microchannel plate detector by a Wiley-McLaren style time-of-flight mass spectrometer \cite{Wiley1955}.

\section{Alternate deceleration method}
\par
To develop efficient slowing protocols, we begin with three-dimensional (3D) Monte-Carlo simulations to understand fully the phase-space dynamics of the deceleration process. Full 3D simulations are required as the separatrix is only a one dimensional representation of the stable phase-space. As pointed out in Refs. \cite{MiejerOptimumAcceptance,mitigation2008}, the transverse velocity and position dimensions modify the phase-space acceptance of the decelerator.
\par
\begin{figure}
\center
\includegraphics[scale=0.8]{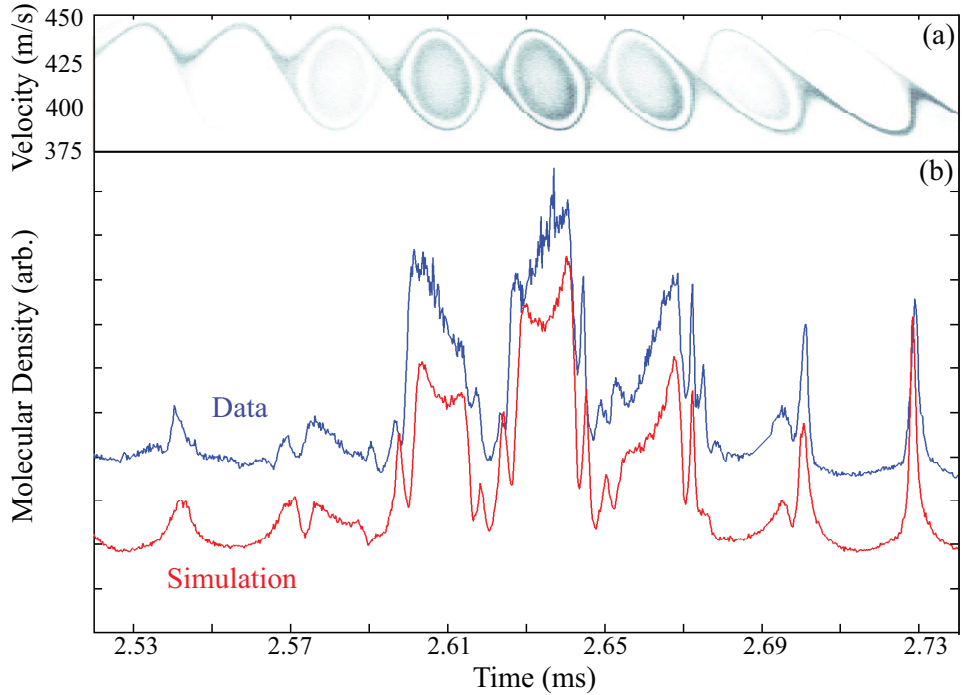}
\caption{Bunching data.(a) Phase-space distribution for bunching ($\phi_0$ = 0$^{\circ}$). (b) A time-of-flight (ToF) trace (blue) is shown with the results of a 3D Monte-Carlo simulation (red). The horizontal axis for both (a) and (b)  is the ToF to the detection laser, which is 1.8 cm past the end of the decelerator. }
\label{bunching}
\end{figure}

This effect can be seen very clearly in our experiment for the case of bunching ($\phi_0 = 0^{\circ}$). Figure \ref{bunching}  shows a time-of-flight profile of the bunched molecular beam including the corresponding simulations. From the plot of the occupied phase space, shown in figure \ref{bunching}(a), one can see the separatrix is not filled uniformly. There are features in the distribution that are considerably more narrow than the characteristic length scale of the decelerator.  Although our ionization-detection scheme essentially integrates this distribution through the velocity dimension, we can still probe this detailed structure experimentally in the position dimension. The structure within experimental signal is elucidated by the simulated phase-space distribution and shows excellent agreement with the 3D simulations. With this agreement, we are confident that our simulations accurately predict the phase-space distribution and thus the energy spread within our slowed molecular packet. This understanding is critical because we can not measure directly the velocity distribution of our molecular beam.

\par
\begin{figure}
\center
\includegraphics[scale=.75]{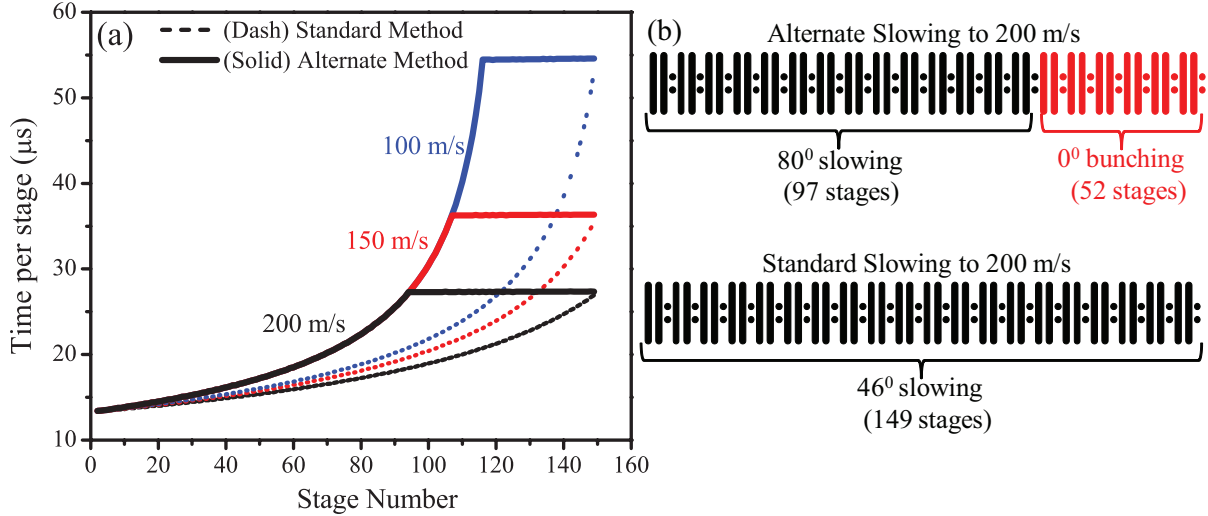}
\caption{ Deceleration schemes using both the standard and alternate methods.  (a) The timing sequence to decelerate a packet of ND$_3$ molecules from 415 m/s to various velocities is shown with the time duration of each stage of deceleration plotted against the stage number.  (b) A schematic representation of the two deceleration schemes decelerating from 415 m/s to 200 m/s.  In the standard method of slowing, all 149 stages of the decelerator are used to decelerate the molecules at a constant 46$^{\circ}$ phase angle.  In the alternate method of slowing, a high phase angle is chosen to aggressively slow the molecules using only the first 97 stages.  The remaining stages are then operated at 0$^{\circ}$ phase angle to longitudinally and transversely guide the molecules to the interaction/detection region. }
\label{timearray}
\end{figure}

The standard method to run a Stark decelerator is to choose a phase angle such that the molecular packet is at the desired velocity when it exits the last stage of the decelerator.  With this method, the energy width of the slowed packet can be relatively large for high mean velocities as it is determined by the phase angle and thus depends on the final velocity. For the method we propose, a large phase angle is chosen ($\phi_0$ = 80$^{\circ}$ in our case), and the number of slowing stages is chosen to decelerate to the final velocity. Once the molecular packet has reached desired velocity, the decelerator is operated in bunching ($\phi_0 =$ 0$^{\circ}$) mode to guide the packet to the end of the decelerator while maintaining the desired velocity. Figure \ref{timearray} illustrates the difference between the standard and alternate methods.  In figure \ref{timearray}(a), the time the molecules spend in a stage of the decelerator as they propagate is plotted for several different final velocities. In the standard method, the molecules are slowed progressively for the entire 149 stages at a different phase angle depending on the desired final velocity. The time spent in a stage increases as the molecules are decelerated. With the alternate slowing technique, the molecules always follow the same trajectory until they reach the desired velocity at which point the switching time of the decelerator remains constant indicating a constant velocity.

\par
\begin{figure}
\center
\includegraphics[scale=.9]{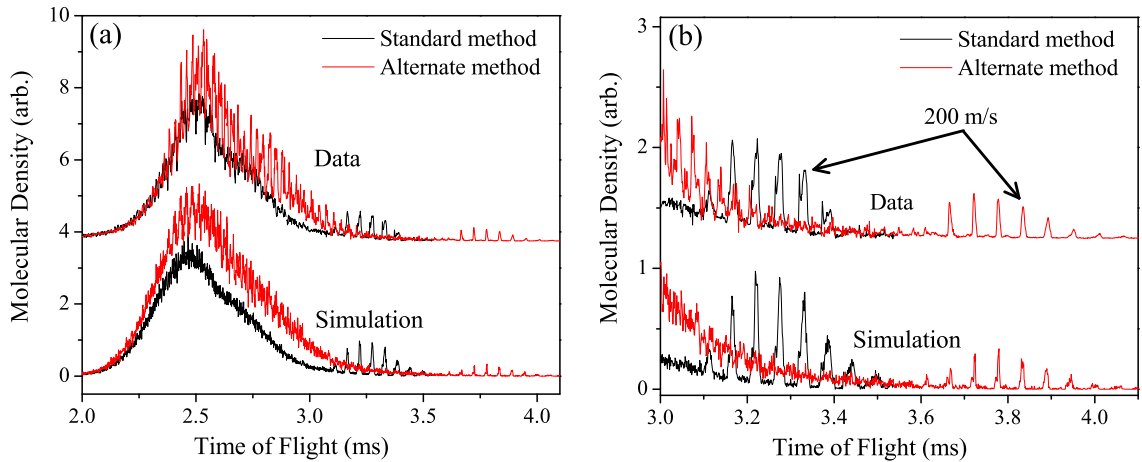}
\caption{ Time-of-flight traces comparing alternate and standard techniques. (a) Time-of-flight  traces showing the arrival time of the decelerated peaks using both the standard and alternate methods of slowing. The traces shown are for a molecular beam that is decelerated from an initial speed of 415 m/s to a final speed of 200 m/s.  Using the alternate deceleration method, the decelerated peak arrives 515 $\mu$s later than the analogous peak using the standard method.  The data for both the standard and alternate methods are taken at 10 Hz, using a 15 mJ laser pulse focused with a 50 cm lens.  Each point is averaged using 10 shots corresponding to a data acquisition of approximately one point/second.  The results from a 3D Monte-Carlo simulation are also shown (lower curves). (b) Expanded time axis shown for clarity.  }
\label{traces}
\end{figure}

One advantage of the alternate slowing protocol is the increased time between the arrival of the slowed packet and the unslowed beam. This is important for two reasons.  First, for relatively large velocities, the slowed packet will ride on a background of the unslowed molecules, thus increasing the energy spread. This effect can be seen in figure \ref{traces}(b) where a packet slowed to 200 m/s with the standard protocol has not separated completely from the unslowed beam. For this case, the unslowed molecules can not be distinguished from the slowed packet. This overlap is undesirable for collision experiments in which collision energy plays a role. Second, it is advantageous to have sufficient time between the unslowed beam and the decelerated molecules to insert another collision reactant into the path. In the case of collisions between trapped atoms or molecules and a slowed beam of molecules, the trapped reactant can be moved into position after the unslowed beam has gone by, thus reducing unwanted collisions with the unslowed beam. An example of this increased separation is shown  in figure \ref{traces}, where we slow a portion of the molecular beam to 200 m/s using both the standard slowing technique and the alternate method. Here we see the arrival time between the unslowed beam and the slowed packet has increased by 0.5 ms.  This increase in time is given by
\begin{equation}
\Delta t=\frac{(V_f-V_i)^2}{2 V_f a_a}+\frac{L}{V_f}-\frac{V_f-V_i}{a_s},
\end{equation}
where $V_i$ and $V_f$ are the initial and final velocities of the packet, respectively, $a_a$ and $a_s$ are the accelerations of the synchronous molecule in the alternate and standard mode, and $L$ is the length of the decelerator. This equation assumes a constant average acceleration during the slowing process that can be determined from the potential created by the electrodes \cite{KoosGubbels2006}.

\par
\begin{figure}
\center
\includegraphics[scale=0.75]{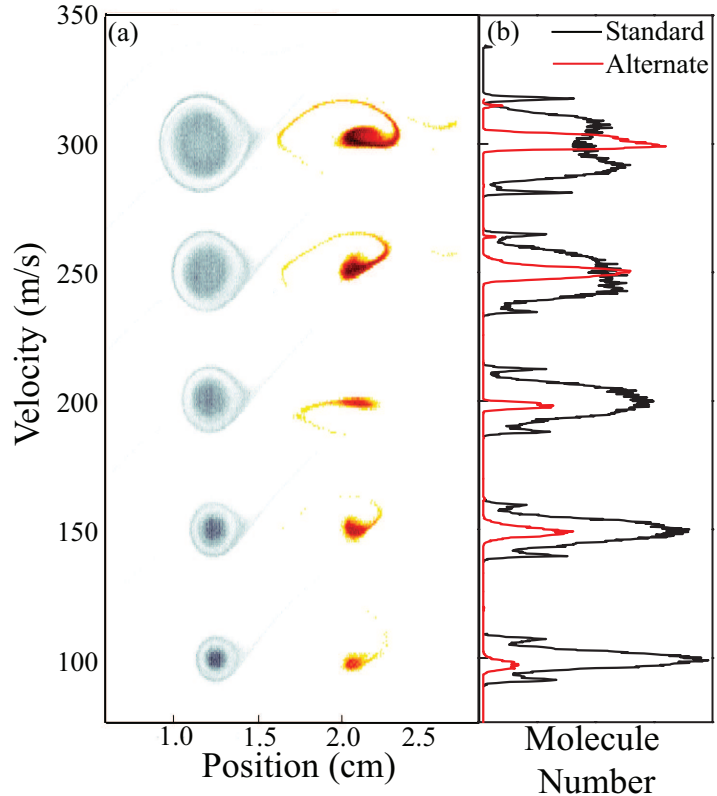}
\caption{ The (a) phase-space distributions at the end of the decelerator with (b) the corresponding velocity histograms.  The distributions for the alternate method (red) are offset from the distributions for the standard method (blue) for clarity. The plots in (b) represent a histogram of velocities taken with a 1 $\times$ 1 mm horizontal cut through the molecular packet in the transverse direction.  }
\label{colormaps}
\end{figure}
\begin{figure}
\center
\includegraphics[scale=0.9]{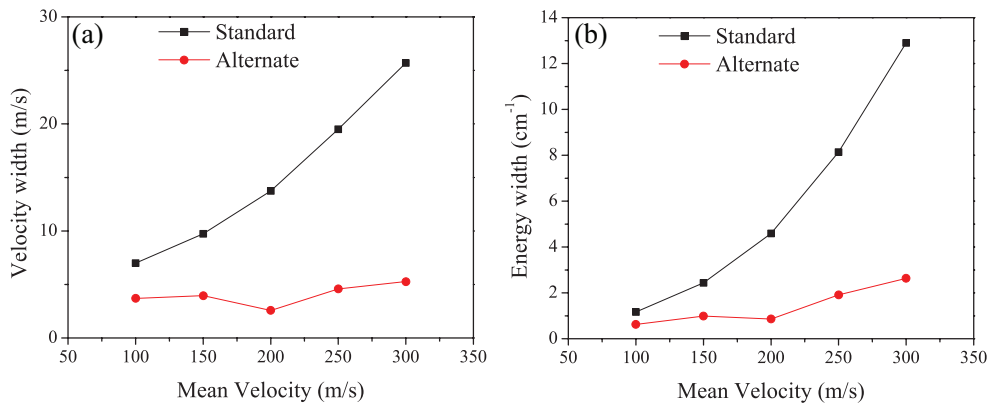}
\caption{The (a) velocity widths and (b) corresponding energy widths of molecular packets for the standard and alternate deceleration methods.  For standard slowing, there is a strong correlation between the phase angle (i.e., final velocity) and the energy width of the decelerated packet.  In the alternate slowing method, however, the same phase angle is used for all final velocities. Therefore there is considerably less correlation, and the velocity width is effectively constant as a function of final velocity.}
\label{widths}
\end{figure}

Another advantage to operating in the alternate mode is the increase in energy resolution for collision experiments. For standard slowing, the velocity spread of the slowed packet increases for increasing velocity, as can be seen in figure \ref{colormaps}. In addition to the large velocity spread, the distribution is far from Gaussian, as expected, leading to difficulty in characterizing the width for extracting parameters of collision measurements. The corresponding histograms for the alternate method are shown in red in figure \ref{colormaps}(b). Here the velocity widths are considerably less, and the distributions are nearer to Gaussians. Because the initial phase-space acceptance is independent of final velocity, the velocity width for the alternate method remains almost constant as the velocity is changed (figure \ref{widths}(a)). For our experimental parameters, we see a reduction in the velocity spread of up to a factor of $\sim$ 5. It is useful to look at the improvement to the spread in terms of energy. The energy width of the packet as a function of final velocity for both methods is shown in figure \ref{widths}(b). The decrease in energy spread of the molecular packet decelerated using the alternate method allows more precise determination of energy-dependent collisional cross-sections. However, this gain comes at a cost to molecular number and thus is not optimal for all types of experiments.

\section{Phase-space rotation}
\par
Realizing a large decrease in energy spread of a molecular packet using our alternate method comes with one caveat. After the molecules are decelerated to the desired velocity, they are transported to the end of the decelerator under bunching conditions. While in the bunching stages, the molecular packet undergoes rotations in phase space. Through this rotation, the phase-space distribution will oscillate between having a wide velocity spread with a narrow position spread and having a narrow velocity spread with a wide position spread. Therefore it is critical to control this rotation such that the phase space has a narrow velocity spread at the detection/interaction region. We accomplish this phase rotation by varying the timing of the bunching in such a way as to change the rate of rotation without affecting the velocity of the packet. Phase-space rotation has also been demonstrated using additional electrodes with a different geometry \cite{PhysRevLett.89.093004}. However, we take advantage of the remaining decelerator stages to obtain the desired rotation.

\par
\begin{figure}
\includegraphics[scale=.8]{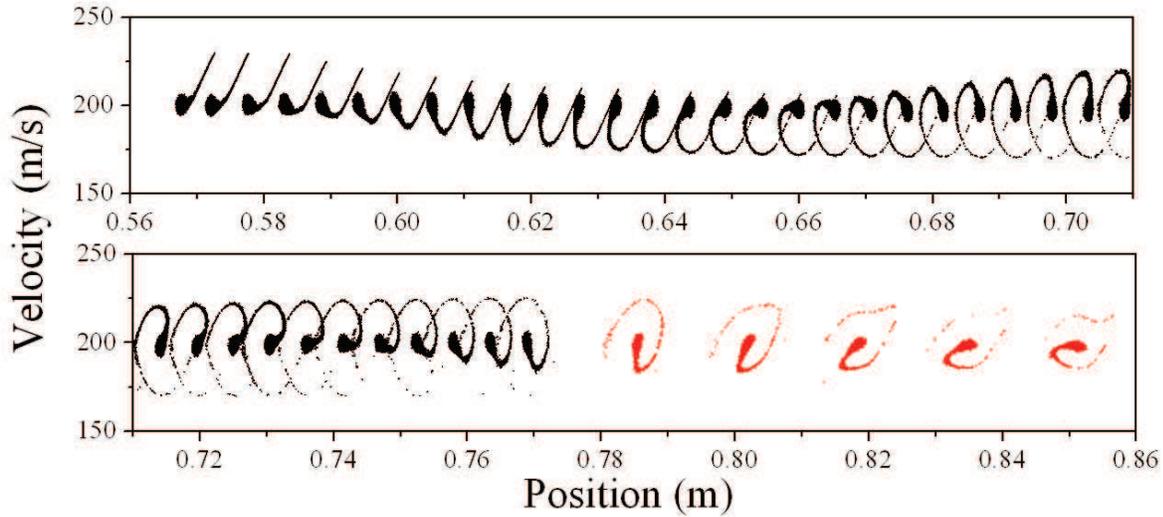}
\caption{ Phase-space rotations of the molecular packet during a bunching sequence. Each image is a snapshot of the phase-space distribution after a single bunching stage. The $S = 1$ stages are shown in black, and the $S = 3$ stages are in red. Oscillations in phase space of the packet during the bunching sequence leads to alternating broadening and narrowing of the velocity distribution as the packet traverses the remainder of the decelerator.  The frequency of the oscillations can be manipulated (and hence the total number of oscillations) by an appropriate application of $S = 1$ and $S = 3$ bunching.  In this example, the molecular packet is decelerated using the first 97 stages of the decelerator to 200 m/s.  The desired rotation is achieved by 37 $S = 1$ bunching stages and 5 $S = 3$ stages.}
\label{rotations}
\end{figure}

Bunching can occur at harmonics of the fundamental bunching frequency while still maintaining phase stability \cite{Sebastiaan2005}. First harmonic bunching occurs when the potential on the electrodes is switched every time the synchronous molecule crosses $\phi_0 = 0^{\circ}$ (i.e., directly between two electrode pairs). For third harmonic bunching, the potential is switched at every third stage. Bunching harmonics can be parameterized by a single parameter, S, which labels the harmonic order. The phase-space rotation frequency depends on the order of the bunching harmonic. The frequency of the rotation to first order is given by
\begin{equation}
\nu_s=\frac{\omega_z}{2\pi}=\sqrt{\frac{\Delta E}{4\pi S m l^2}},
\end{equation}
where $\Delta E$ is the maximum possible amount of Stark energy lost in a stage, $m$ is the mass of the molecule, and $l$ is the separation between the electrodes in the decelerator \cite{Sebastiaan2005}.  Under our experimental conditions, the frequency of rotation for $S = 1$ bunching is $\sim$1.4 kHz.   Using both $S = 1$ and $S = 3$ bunching harmonics, we can manipulate the amount of rotation.  The number of rotations the packet will undergo is given by
\begin{equation}
N_{rot}=\nu_1 t_1\left[n_1\left(1-\frac{1}{\sqrt{3}}\right)+\frac{n_T}{\sqrt{3}}\right],
\end{equation}
where $t_1$ is the time for an $S = 1$ bunching stage, $n_1$ is the number of $S = 1$ stages, and $n_T$ is the total number of physical stages used for bunching.  Figure \ref{rotations} shows the rotation in phase space of a packet decelerated to 200 m/s as it is bunched through the remainder of the slower. In this case, we use 37 $S = 1$ stages and 5 $S = 3$ stages, which results in a narrow velocity spread. The phase-space distribution undergoes just over 1.5 rotations and thus maintains the narrow velocity width created by our alternate slowing method.  Figure \ref{colormaps} shows a tail of phase-stable molecules rotating around the packet in phase space.  Although these molecules have a different velocity from the main packet, they contribute only a very small fraction to the total number of detected molecules, typically on the order of one percent.

\par
\begin{figure}[!h]
\center
\includegraphics[scale=.7]{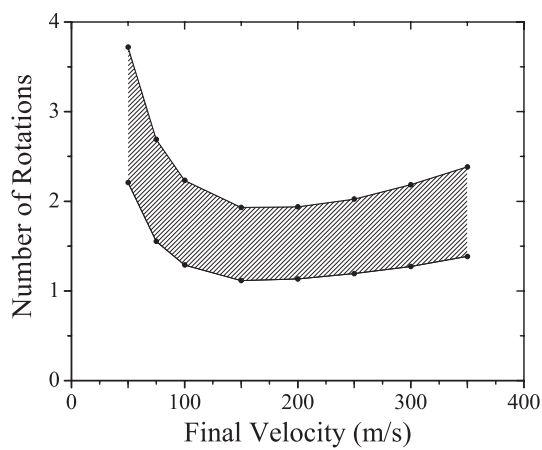}
\caption{ Calculated range over which the total number of rotations can be manipulated as a function of the final velocity for a decelerator of fixed length. The shaded region represents the accessible rotations.}
\label{rotationrange}
\end{figure}

There is a finite range of rotations accessible with a particular decelerator design and final velocity. The maximum number of rotations is achieved by using only $S = 1$ stages, and the minimum is achieved by using only $S = 3$ stages. This range is shown in figure \ref{rotationrange}. For most velocities, we can access any rotation angle.

In conclusion, we have developed a new protocol to run a Stark decelerator to achieve narrow energy spreads over a large range in velocities. We reduce the energy spreads by up to a factor of approximately five over traditional methods. These high-energy-resolution beams will enhance a variety of cold-molecule collision experiments using Stark decelerators.
\ack
This work was supported by NSF, AFOSR, Petroleum Research Fund, and the Alfred P. Sloan Foundation.

\section*{References}


\end{document}